\documentstyle[aps,prb]{revtex}

\begin{document}

\twocolumn[\hsize\textwidth\columnwidth\hsize\csname
@twocolumnfalse\endcsname

\draft
\author{Horacio M. Pastawski$^{+*}$ and Gonzalo Usaj$^{+}$}
\address{$^{+}$Facultad de Matem\'{a}tica, Astronom\'{\i}a y F\'{\i}sica. Universidad Nacional 
de C\'{o}rdoba. Ciudad Universitaria. 5000 C\'{o}rdoba. Argentina\\
$^{*}$International Center for Theoretical Physics, P.O. Box 586. 34100}
\title{Dimensional Crossover in Spin Diffusion: A manifestation of the Quantum Zeno Effect }
\date{Accepted in PRB, Scheduled tentatively B1 01Mar98 }
\maketitle

\begin{abstract}
The Quantum Zeno Effect (QZE) implies that a too frequent ($\omega _\phi
\rightarrow \infty )$ observation of a quantum system would trap it in its
initial state, even though it would be able to evolve to some other state if
not observed. In our scheme, interacting spins in a 3-d cubic lattice,
``observe'' each other with a frequency $\omega _\phi $ $\propto \sqrt{J_x^2+J_y^2+J_z^2}/\hbar $, where $J$'s are the coupling constants. This
leads to a ``diffusive'' spread of a local excitation characterized by the
constants $D_\mu \propto J_\mu ^2/\omega _\phi .$ Thus, a strongly
asymmetric interaction (e.g. $J_y/J_{x(z)}\gg 1$), would hinder diffusion in
the perpendicular directions ($D_{x(z)}\rightarrow 0$) manifesting the QZE.
We show that this effect is present in numerical solutions of simple 2-d
systems. This reduction in the diffusion kinetics was experimentally
observed in paramagnetic compounds where the asymmetry of the interaction
network manifests through an exchange narrowed linewidth$.$ New experimental
designs are proposed.
\end{abstract}

\vskip2pc]\narrowtext

Quantum dynamics of magnetic excitations in a system of interacting spins at
high temperature is an active field of research\cite{Stolze}. From a
macroscopic scope, i.e. for long times and large wave-lengths, one expects
that excitations should evolve irreversibly. The resulting hydrodynamic
equations\cite{Forster} describe the\ ``spin diffusion''. In this regime a
local excitation decay as $(Dt)^{-d/2}$ where $d$ is the dimension of the
space. From the microscopic point of view, however, dynamics is governed by
reversible quantum mechanics as long as quantum coherence is maintained.
Although this dynamics {\it seems} to be ``diffusive'' already at
intermediate times, a careful study\cite{PLU} of low dimensional systems can
recognize quantum interferences. Its experimental observation\cite{PUL,Madi}
constitutes a fingerprint of the bounded regions where the dynamics occurs.
Besides, an experimental realization of a ``Loschmidt daemon''\cite{daemon},
which allows the evolution backwards in time, can be achieved by inverting
the sign of the effective Hamiltonian. In this case, ``irreversible''
interactions are simply those we do not control. Hence, reversibility is not
total\cite{ZME}. This effect is stronger than what can be inferred from the
magnitude of the non-inverted terms. In fact, the apparently ``diffusive''
dynamics of the many body interaction seems to transform small residual
interactions into efficient mechanisms to stabilize an irreversible diffusion%
\cite{LUP}. Irreversible effect is often amplified even by those
interactions we {\it can} control. This is analogous to the Drude
approximation for the electrical resistance of an impure metal at low
temperatures. There, the {\it reversible} elastic scattering with impurities
with rate ($1/\tau _{{\rm imp}}$) facilitates the interaction with the
thermal bath. This bath acts through the uncontrollable electron-phonon
interactions or other dephasing collisions with rate $1/\tau _{{\rm \phi }}$%
. The remarkable consequence is that in first approximation the diffusion
constant does not depend on $\tau _{{\rm \phi }}$ but it is $D\propto
v_{{}}^{2}\tau _{{\rm imp}}$, where $v$ is a typical velocity for ballistic
propagation of the excitation. Its lesson is that a reversible interaction
can provide an evolution close enough to diffusion. Then, when actual
irreversible processes occur, they stabilize\cite{GLBE} the evolution into
an irreversible diffusion. The observed diffusion constant remains the
unchanged. This gives the ultimate justification for the {\it %
stosszahlansatz }or assumption of randomness after successive collisions in
which no memory of the previous quantum state is retained. This is
equivalent to consider a collision with an impurity as a classical
measurement or wave function collapse. This, of course, is an approximation
which breaks down close to the localized regime where interferences play a
fundamental role.

The goal of the many-body techniques is to provide the match between the
quantum and hydrodynamic regimes providing the machinery for the calculation
of the diffusion constant and other macroscopic observables. However,
because of the implicit approximations, this is sometimes done at the cost
of the physical insights into the nature of irreversibility. To understand
the dynamics of a local excitation, we notice that one spin up, in a lattice
with all spins down, propagates with a typical ``ballistic'' velocity
proportional to the exchange constant $J$. At high temperature, there are as
many up spins as down ones. Hence, this propagation in the sublattice of
spins down is interrupted by the modification of this lattice. This produces
a ``collision'' rate ($1/\tau _{{\rm mb}}$) also proportional to $J.$ If we
neglect quantum interferences ({\it stosszahlansatz}) produced by multiple
collisions, a diffusion constant $D\propto v_{{}}^{2}\tau _{{\rm mb}}\propto
J$ is obtained$.$ Again the point is that a diffusive behavior is a good
approximation for the dynamics. It will be stabilized by later irreversible
interactions. Keeping this in mind, we may {\it consider} each ``{\it %
collision}'' {\it as} {\it a} {\it measurement} process.

Quantum dynamics is strongly modified by recurrent measurements. This
phenomenon, known as the {\it Quantum Zeno Effect} (QZE)\cite{QZeno}, has
been applied with particular success in quantum optics\cite{Interaction-free}%
. In simple words, it affirms that {\it if the evolution} of a quantum
system {\it is observed too frequently}, {\it there will be no evolution} to
be seen! While this might sound paradoxical from a classical point of view,
within the quantum logic this is not a paradox at all, since a measurement
involves a collapse of the wave function and the start of a new quantum
evolution. For short times, the probability of staying in the initial state
is $P_{i,i}(t)=1-(\overline{J}t/2\hbar )^{2}+\ldots ,$ with $\overline{J}$
being an average exchange energy. The lack of a linear term on this
expansion has very important consequences. If the evolution in a time $t_{s}$
is interrupted by $N$ observations, the final probability of stay is $%
\widetilde{P}_{i,i}(t=t_{s})$ =$\left[ P_{i,i}(t_{s}/N)\right]
^{N}\rightarrow 1$ when $N\rightarrow \infty $. Here, $t_{s}/N=\tau _{\phi }$
defines the decoherence time. Hence one sees why a frequent collapse ( i.e. $%
\omega _{\phi }=1/\tau _{\phi }\gg 1$) hinders evolution\cite{LPDA}. The
result of the successively interrupted evolution is shown schematically in
Fig. 1.

This article addresses a paradoxical aspect of the dimensional dependence in
the dynamics of magnetic excitations at the light of the QZE. In order to
make it more obvious, let us firstly discuss the particular limit aa cubic
lattice where the interaction along direction $y$ grows ($%
J_{y}/J_{x(z)}\rightarrow \infty $). The diffusion coefficient in that
direction will also grow: $D_{y}\propto J_{y}.$ However, the spreading rate
in the other directions will decrease $D_{x,(z)}\propto 1/J_{y}.$ The
essence of the argument that we are going to develop is that spins spreading
in each $xz$-plane register the evolution in the other parallel planes
through the many body coupling $J_{y}$ (i.e. they ``observe'' each other
with a frequency $\omega _{\phi }\rightarrow J_{y}/\hbar $ ), producing the
reduction of the dynamics within the planes. This is a general behaviour
also valid for the non-pertuvative limit of $J_{y}\approx J_{x(z)}.$ To our
knowledge this has not been noticed previously.

The calculation of the dynamics of a local spin polarization in the high
temperature regime can be mapped to a system of Fermi particles on a
lattice. Up spins are identified with particles and down spins with holes. A
particle is localized at the excited site with the rest of the sites being
occupied with probability one half. The initial state is an incoherent
superposition of all the possible initial states a few of which are shown in
the inset of Fig. 1. The particle initially at 0th site starts to evolve
having a finite probability amplitude to jump into empty neighboring sites.
Meanwhile, other particles also can move in and out of neighboring sites
creating a fluctuating effective potential. If this is approximated by a
stochastic potential, one obtains an irreversible equation satisfying the
hydrodynamic limits. The essential point is that the correlation times of
the potential are the same that characterize the dynamics of a particle in
its fluctuating environment. This leads to a {\it self-consistent} equation
which is the core of classic many-body calculations of spin ``diffusion''
such as that of Blume and Hubbard\cite{Blume} for symmetric lattices.
However, those calculations do not show the mind-teasing behavior of the
asymmetric lattices which are the purpose of this work. Let us do a
simplified calculation valid for both symmetric and asymmetric systems,
which will clarify the effect of dimensional crossover in the dynamics.
Consider a $d$-dimensional (hyper)-cubic lattice of spins 1/2 interacting
through a Hamiltonian:

\begin{equation}
{\cal H}_{II}=\sum_{k,j>k}J_{jk}\,\left[ \alpha 2S_j^zS_k^z-\frac 12\left(
S_j^{+}S_k^{-}+S_j^{-}S_k^{+}\,\right) \right] ,  \label{Hii}
\end{equation}
where $S$ are the usual spin operators with subscripts indicating spin sites
and $J_{jk}=J_\mu $ are nearest neighbors interaction parameters depending
only upon the direction $\widehat{\mu }$ along sites $j$ and $k$ at distance 
$a_\mu \equiv a=1$. For $\alpha =-\frac 12$, it describes the Heisenberg
model (isotropic exchange), $\alpha =0$ defines the XY model, while $\alpha
=1$ is a truncated dipolar Hamiltonian.

For each initial state$,\mid i\rangle ,\,$with $0$-th site polarized (i.e.
one of the states shown in Fig. 1), the probability of finding the same site
polarized in the state $\langle f\mid $ after a time $t,$ is 
\begin{equation}
P_{f,i}(t)=\left| \langle f\mid \exp [-\frac{{\rm i}}\hbar {\cal H}%
_{II}{}t]\mid i\rangle \right| ^2.  \label{Pfi}
\end{equation}
A total ensemble averaged probability that a spin initially up at position $%
0 $-th is still up at time $t$ can be calculated summing over all the $N_i$
and $N_f$ possible initial and final states:

\begin{eqnarray}
\left\langle P(t)\right\rangle &=&\sum_f^{N_f}\sum_i^{N_i}\frac 1{N_i}%
P_{f,i}(t)  \nonumber  \label{Pt} \\
&=&1-\frac 14\sum_\mu ^d\left\langle Z_\mu \right\rangle J_\mu ^2t^2/\hbar
^2+{\cal O}(t^4)+\ldots .  \label{Pt}
\end{eqnarray}
$\left\langle Z_\mu \right\rangle =1$ is the average number of neighbors
along direction $\mu $ with spin down. In the second order term only the
flip--flop terms produce the exchange of the originally polarized $0$-th
spin with its neighbors. This corresponds to the {\it one body} dynamics of
one up spin in a lattice in which the other up spins remain frozen. Higher
order terms contain the dynamics of those other spins and the {\it many body}
interactions.

A normalized magnetization can be calculated from the spin autocorrelation
function as 
\begin{equation}
M(t)=\left\langle S_{0}^{z}(t)S_{0}^{z}\right\rangle /\left\langle
S_{0}^{z}S_{0}^{z}\right\rangle =2(\left\langle P(t)\right\rangle -\frac{1}{2%
}).  \label{M}
\end{equation}
This magnitude is experimentally accessible. For short times, the mean
square displacement of the magnetization in terms of the nearest neighbor
spins correlation functions: $\left\langle r_{\mu }^{2}\right\rangle \propto
a_{\mu }^{2}\left\langle S_{0\pm a_{\mu }}^{z}(t)S_{0}^{z}\right\rangle .$
The truncated quantum dynamics given by Eq. (\ref{Pt}) gives :

\begin{equation}
\left\langle r_{\mu }^{2}\right\rangle _{quant}=\frac{1}{2}a_{\mu
}^{2}J_{\mu }^{2}t^{2}/\hbar ^{2}.  \label{x2q}
\end{equation}
However, the perturbative expansion of Eq.(\ref{Pt}) is not a practical way
to obtain the long time dynamics but for few simple one dimensional systems%
\cite{Stolze}. Therefore, we need an entirely different approach. First, we
notice that for infinite lattices the long time evolution is very complex
and it presents the {\it apparently} diffusive behavior (eventually
stabilized by interactions with the thermal bath) that we want to evaluate.
While evolution with (\ref{Hii}) is not an irreversible process, it is close
enough to a diffusive evolution so that a self-consistent condition is
already achieved for intermediate times.

When $\left\langle P(t)\right\rangle $ has decayed substantially, let us say
when the second term in Eq.(\ref{Pt}) is half the first, the environment has
changed completely and no phase coherence with the initial {\it one-body}
state is retained. This defines the dephasing time

\begin{equation}
\tau _\phi =\hbar 2/\sqrt{2\sum_\mu ^dJ_\mu ^2}=1/\omega _\phi .
\label{1/tau}
\end{equation}
It is important to note that this functional dependence on $J_\mu $ does not
depend on the amount of decay chosen to determine $\tau _\phi $. The change
in the environment, following itself a quantum dynamics according to Eq. (%
\ref{Pt})$,$ is slower than linear at early times, becoming important only
at about $\tau _\phi .\,$ Then, it can be described by a discrete time
Markovian process leading to a classical random walk. At the dephasing time
the mean square displacement of the magnetization is 
\begin{equation}
\left\langle r_\mu ^2\right\rangle _{class.}=2D_\mu \tau _\phi ^{}.
\label{x2c}
\end{equation}
At $t=\tau _\phi ,$ both quantum and classical diffusive regimes must
coincide. This statement was rigorously proved in Ref.\cite{GLBE}, where we
used the Keldysh's formalism\cite{GLBE2} to achieve a {\it non-perturbative}
description of the cross-over from the quantum to the diffusive regime for a
particle interacting with a dephasing field. In our model this field acts at
a typical time with an interaction probability $p(t){\rm d}t=\delta (t-\tau
_\phi ^{}){\rm d}t$. Then, the coherence with the initial state has a
survival probability $\theta (\tau _\phi ^{}-t).$ This step function is more
appropriate to describe the quantum dynamics of the dephasing field than the
usual $\exp [-t/\tau _\phi ^{}]$. According to Ref.\cite{GLBE}, within this
approximation the {\it self-consistent }propagation of density excitations
(satisfying the integral Keldysh equation) requires that both quantum and
Markovian descriptions\cite{LPDA} give the same probability distribution at $%
t=\tau _\phi ^{}.$ This condition is equivalent to equate (\ref{x2q}) and (%
\ref{x2c}), from which we obtain the diffusion constant for each direction:

\begin{equation}
D_{\mu }=\frac{a_{\mu }}{2\hbar }\times J_{\mu }^{2}/\sqrt{2\sum_{\mu
}^{d}J_{\mu }^{2}}=\frac{a_{\mu }^{2}}{4\hbar ^{2}}J_{\mu }^{2}/\omega
_{\phi }  \label{Di}
\end{equation}
This important result contains the paradoxical aspects of spin dynamics we
discussed in the introductory paragraphs. While our procedure has been
mainly qualitative, we believe it catches the fundamental phenomena, and
therefore the correct functional dependence. For the symmetric three
dimensional lattice this gives $D=Ja^{2}/(\hbar 2\sqrt{6})$, in fair
agreement with the values calculated by a number of previous authors\cite
{Blume,SpinDiff} and consistent with simulations\cite{Blume} in a system of
classical spins.

While numerical solution of the quantum dynamics of systems with a large
number of spins is a formidable task, we can attempt to see signatures of
the discussed phenomena for small two dimensional systems. We consider a
nine spins system with periodic boundary conditions and evaluate the
eigenstates of all the spin configurations to study the dynamics according
to Eqs. (\ref{Hii}) and (\ref{Pfi}) with $\alpha =-1/2$. In Fig.2 we show
the numerical evaluation of $\left\langle r_x^2\right\rangle $ and $%
\left\langle r_y^2\right\rangle $ as a function of time $t$ for different
values of $J_y$ while keeping $J_x=1.$ $.$ While the diffusive regime $%
(\left\langle r_y^2\right\rangle \sim 2D_yt$) can not be reached in a small
system, the plot shows that an increase in $J_y$ is correlated with an
increase in the spreading dynamics. The paradox that manifest the QZE is
that the growth of $\left\langle r_x^2\right\rangle ,$ showing the dynamics
in the perpendicular direction, is slowed down by the increase of $J_y$.
While eventually the hydrodynamic limit could be described by a diffusion
equation in which variables can be separated, in the many body
Schr\"{o}dinger equation variables appear intimately entangled, leading to
the interdependence of the diffusion constants of Eq. (\ref{Di}).

We want to show that even when an asymmetry of the lattice could lead to a
faster quantum decay of the polarization for short times, it produces a
reduction of the effective dimensionality of the lattice where diffusion
occurs which slows down the spreading. For this purpose we study the time
decay of the local polarization in a square lattice. The inset in Fig. 3
shows the magnetization $M(t)$ in the symmetric lattice up to intermediate
times (for long times weak mesoscopic beats\cite{PUL} resembling to those of
1-d rings would appear). The symmetry of the interaction network can be
broken by slightly increasing the coupling along direction $y$, $%
J_y=J+\delta J,$ while in the other direction it is decreased by the same
amount $J_x=J-\delta J$. The thick line in the main plot shows the
difference between the local magnetizations calculated for the asymmetric
and the symmetric networks $\delta C=M^A(t)-M^S(t),$ with $\delta J/J=0.1.$
For very short times the spreading in the asymmetric lattice is faster than
in the symmetric one and $\delta C$ follows the parabolic approximation $%
\delta C\approx -(\delta J)^2t^2/\hbar ^2$ shown by the thin line. QZE
manifests for intermediate times, slowing down the diffusion in the
asymmetric case. This compensates the fast decay of short times at around $%
t=1.8\hbar /J$, when $\delta C=0.$ This crossover from faster decay of $M^A$
as compared with $M^S$ to a slower one is a very remarkable result of our
simple theory consistent with the numerical solutions.

A simple experimental test of the spin dynamics is the linewidth of a
magnetic resonance spectrum. Different local environments in a system of
non-interacting spins produce an inhomogeneously broadened absorption line,
where each frequency corresponds to spins seeing a different local field.
However, through the flip-flop mechanism, each spin excitation explores
different lattice sites producing an averaging of the field. Thus, the
narrowed linewidth \cite{exchange-narrow} is: $\Delta \nu \propto
\int_0^\infty \left\langle S_0^z(t)S_0^z\right\rangle {\rm d}t$ (eventually
the integral extends up to a cut-off time). Then, the slower the dynamics
the wider the line.

It is not usually possible the control of the magnitude of the exchange
interaction for electronic spins. However, this is just the situation
observed\cite{Cu-aa} for a family of Cu(aa)$_2$ single crystals where aa
stands for amino acid. If the aa is present in a mixture 50\% dextrogyre and
50\% levogyre (D,L), the Cu$^{2+}$ paramagnetic centers occupy sites with
inversion symmetry in a two-dimensional coupling network. The Cu atoms are
connected by OCO bridges, with the different aa residues playing the role of
separators between layers. In Cu(L-aa)$_2$ crystals (100\% L-aa), however, a
breakdown of this symmetry occurs. The amount of asymmetrization of the
lattice depends on the aa. The very important increase of the electron
paramagnetic resonance linewidth that is observed when going from D,L to L
crystals can be interpreted as a signature of the discussed QZE.

Systems with interacting nuclear spins seem quite promising for the study of
dynamics in asymmetric lattices. Macro-molecules could be engineered to
present a sequence of through bond isotropic couplings when studied in
solution. In crystals, where the dipolar interaction is the dominant one,
its dependence on the angle $\theta $ between the vector connecting dipoles
and dipole orientation (fixed by an external field) can be used to change
its magnitude and sign. For example, in a cubic lattice of spins, by varying
the magnetic field in the plane [001] from the [110] direction toward one at
a magic angle ($\theta _m=\arccos [1/\sqrt{3}]$) with the [010] axis, one
could see how the dynamics changes from a three-dimensional behavior ($%
J_x=-d/2$, $J_y=-d/2$, $J_z=d$), with a symmetric diffusion in the $xy$%
-plane, toward that of a two dimensional system ($J_x=-d$, $J_y=0$, $J_z=d$%
). The change from the initial orientation would allow a study of a
crossover in $\delta C$ similar to that discussed above. The application of
these concepts would be even more direct for magnetically two-dimensional
systems ($J_z\approx 0$) where $J_x$ and $J_y$ can be controlled
independently. In both cases multiple quantum coherence\cite{MQC}
experiments and spin diffusion pulse sequences\cite{ZME} could be used to
obtain complementary information\cite{Tomaselli} about the asymmetric
spreading of magnetic polarization. A more conventional experiment is to
apply a magnetic field gradient along one crystal direction (say $y$) which,
by detuning the resonance frequencies of nuclei at different $xz$ planes, is
equivalent to a decrease of the inter-plane coupling. Results obtained using
this technique\cite{Cory} are also consistent with our predictions. In
summary, we have put the problem of spin diffusion under a new perspective
that could stimulate new series of experiments and calculations to
understand spin dynamics.

This work was done at LANAIS de RMN (UNC-CONICET) with financial support
from Fundaci\'{o}n Antorchas, CONICET, CONICOR and SeCyT-UNC. We are
grateful to P. R. Levstein for critical reading of this manuscript and
numerous discussions. HMP also acknowledges M. Tomaselli and Prof. D. Cory
for conversations about their unpublished results.

{\bf Figure 1.} Probability of stay for a particle whose quantum evolution
is interrupted by a measurement process with a period $\tau _{\phi }$ and $%
\tau _{\phi }^{\prime }=2\tau _{\phi }.$ The coherent evolution is
approximated up to the quadratic term and the return probability is
neglected. The increase of the probability of stay is a manifestation of the
QZE. The evolution shown is equivalent to the average evolution of the
ensemble in the inset. Only 5 of the 16 configurations corresponding to a
local excitation in the high temperature limit are sketched.

{\bf Figure 2.} Mean square displacement of magnetization as function of
time for increasing asymmetrization of the 2-d lattice with Heisenberg
interaction. The full line is $J_y$=$J_x$ =$J,$ the other lines are the
sequence $J_y/J$=2$,$ 3, 4, 5 and 6. $J_x$ is kept constant. Upper curves
are displacements along $y,$ while lower ones are along $x.$ The QZE
manifests in the slow down of the spreading along $x$.

{\bf Figure 3}. The inset shows the spin autocorrelation function in a
symmetric 2-d lattice with Heisenberg interaction as function of time. Thick
curve in the main frame shows the difference, $\delta C$, between the
autocorrelation function of the symmetric ($J_y$=$J_x$ =$J)$ and asymmetric (%
$J_y=J+\delta J$ and $J_x=J-\delta J,$ with $\delta J/J=0.1)$ lattices$.$
The thin line is a parabolic approximation.

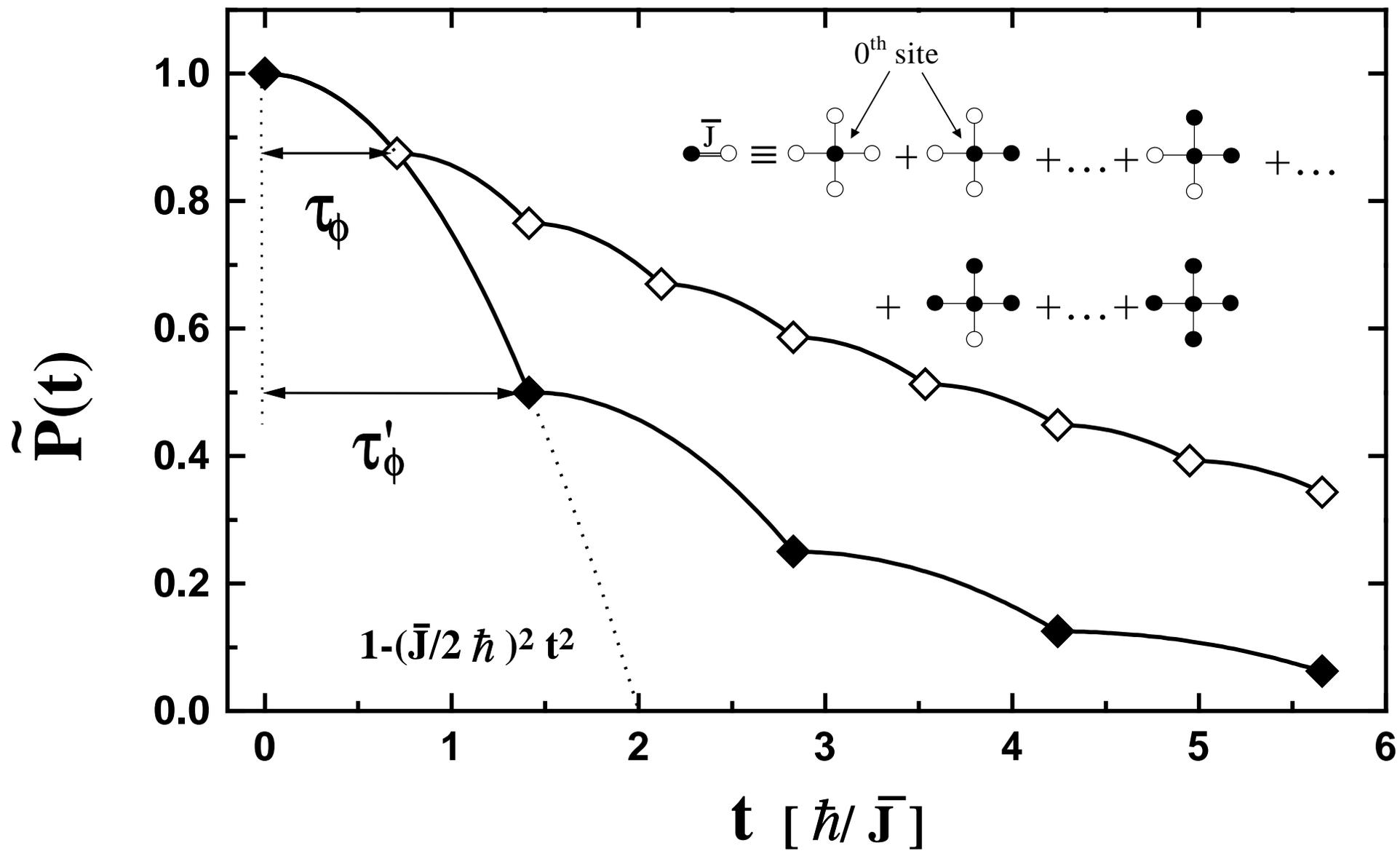
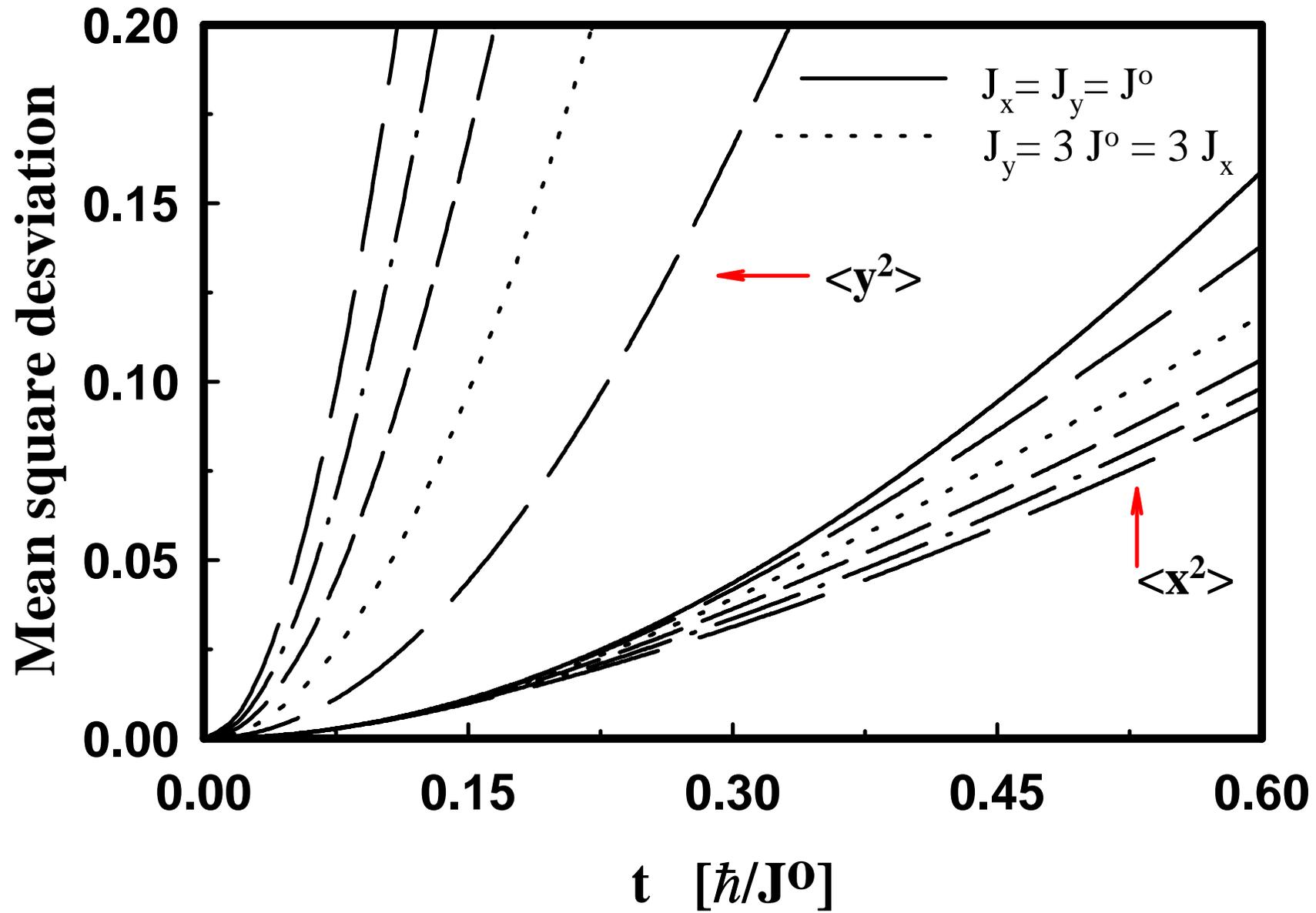
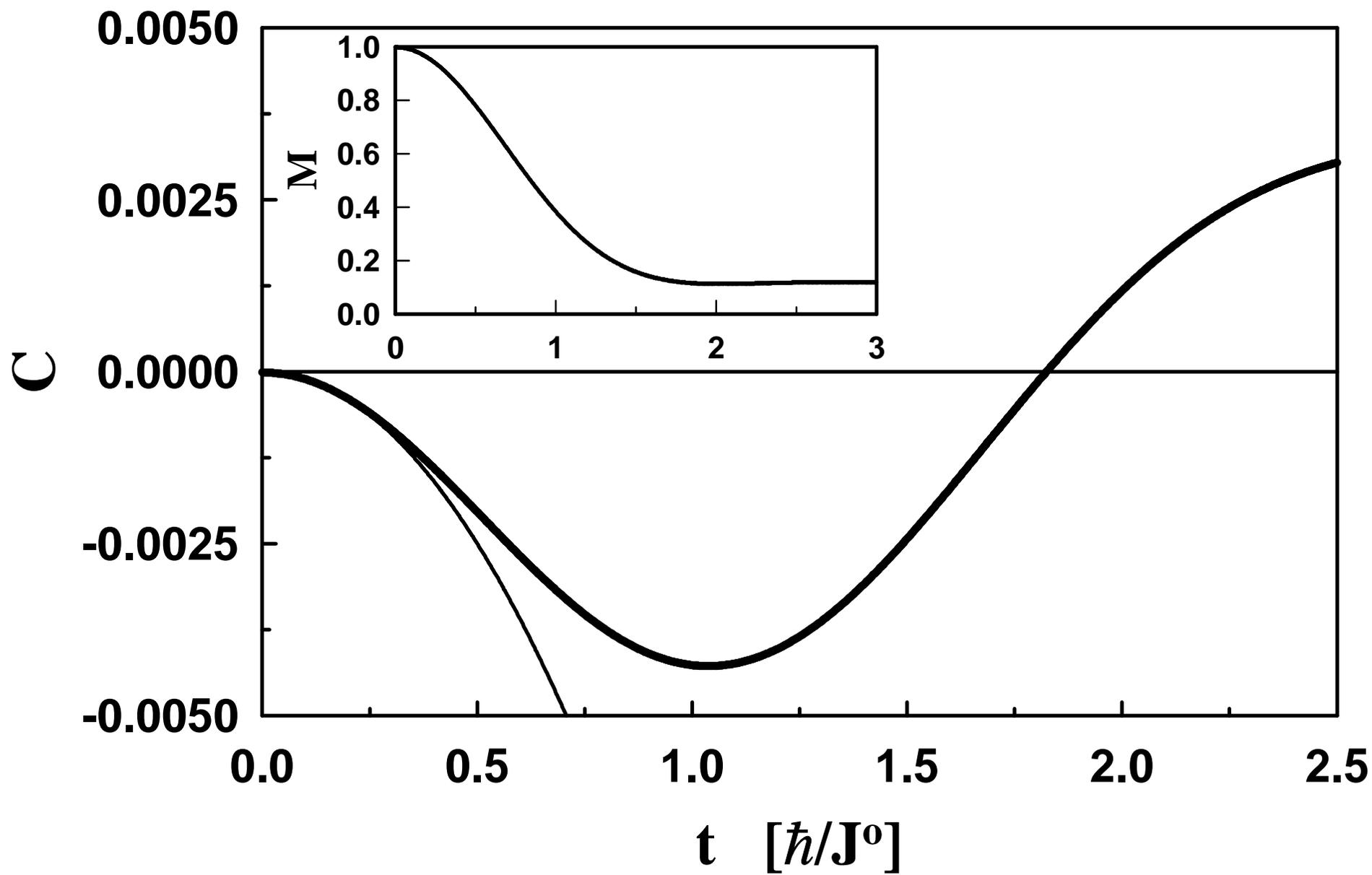
\end{document}